\documentclass[iop]{emulateapj}
\usepackage{graphicx}
\usepackage{natbib}
\usepackage{amssymb}
\usepackage{amsmath}
\usepackage{color}
\bibpunct[; ]{(}{)}{,}{a}{}{;}

\newcommand{\appropto}{\mathrel{\vcenter{
  \offinterlineskip\halign{\hfil$##$\cr
    \propto\cr\noalign{\kern2pt}\sim\cr\noalign{\kern-2pt}}}}}

\shorttitle{Dust Formation Timescales in Novae}
\shortauthors{Williams et al.}

\begin{document}

\title{Rapid Dust Formation in Novae:\\ The Speed Class - Formation Timescale Correlation Explained}

\author{S.~C. Williams, M.~F. Bode and M.~J. Darnley}
\affil{Astrophysics Research Institute, Liverpool John Moores University, IC2, Liverpool Science Park, 146 Brownlow Hill, Liverpool, L3~5RF, UK}
\email{S.C.Williams@2010.ljmu.ac.uk}

\author{A. Evans}
\affil{Astrophysics Group, Lennard Jones Laboratory, Keele University, Keele, Staffordshire, ST5~5BG, UK}

\author{V. Zubko}
\affil{Goddard Space Flight Center, 8800 Greenbelt Rd., Greenbelt, MD~20771, USA}

\and

\author{A.~W. Shafter}
\affil{Department of Astronomy, San Diego State University, San Diego, CA 92182, USA}

\begin{abstract}
Observations show that the time of onset of dust formation in classical novae depends strongly on their speed class, with dust typically taking longer to form in slower novae. Using empirical relationships between speed class, luminosity and ejection velocity, it can be shown that dust formation timescale is expected to be essentially independent of speed class. However, following a nova outburst the spectrum of the central hot source evolves, with an increasing proportion of the radiation being emitted short-ward of the Lyman limit. The rate at which the spectrum evolves also depends on the speed class. We have therefore refined the simple model by assuming photons at energies higher than the Lyman limit are absorbed by neutral hydrogen gas internal to the dust formation sites, therefore preventing these photons reaching the nucleation sites. With this refinement the dust formation timescale is theoretically dependent on speed class and the results of our theoretical modification agree well with the observational data. We consider two types of carbon-based dust, graphite and amorphous carbon, with both types producing similar relationships. Our results can be used to predict when dust will form in a nova of a given speed class and hence when observations should optimally be taken to detect the onset of dust formation.
\end{abstract}

\keywords{novae, cataclysmic variables}

\section{Introduction}

Classical nova outbursts occur in binary systems where a white dwarf accretes matter from a less evolved secondary (either main sequence, sub-giant or red giant; see \citealp{dar12} for a discussion). The outburst is produced when the white dwarf has accreted sufficient matter for the material at the base of the accreted envelope to have a pressure and temperature high enough for nuclear fusion to occur. This leads to a thermonuclear runaway, causing the rapid rise in luminosity and ejection of accreted material observed during outburst (see e.g.\ \citealp{bod10} and references therein).

Dust formation has been observed in Galactic novae for about 40 years (\citealp{gei70}; see \citealp{bodeva89} for a review of the early observations), and observations have recently extended to the potentially much larger sample of extragalactic novae (\citealp{sha11}). A nova outburst in a system containing a CO white dwarf (CO nova) can produce silicates, silicon carbide, carbon or hydrocarbons, or indeed a combination of these (\citealp{geh98,ege12}). Only a small portion (probably $\lesssim0.3\%$; \citealp{geh98}) of a galaxy's interstellar dust originates in novae, but novae may be an important source of some elements and isotopes; indeed they are predicted to be the major source of $^{13}$C, $^{15}$N and $^{17}$O \citep{sta72,josher98,sta08b}.

The link between dust formation and speed class was first suggested by \citet{1977AJ.....82..209G}. \citet{bod82} explored the relationship between speed class and IR development to explain why fast novae seemed less able to form extensive dust shells. We now know that it is the generally very fast Neon novae, however, that produce little or no detected dust. It may be that CO novae generally produce more dust because they contain lower mass white dwarfs, hence an outburst involves higher ejected mass at lower velocities \citep{geh98}. Any dust formed in the nova environment is subjected to the strong radiation field that it produces. However some factors of the environment in nova ejecta are favourable to dust formation, such as the high densities and heavy element abundances found therein. The rates of many reactions which must take place in order for nucleation sites to form are also higher at the temperatures found here \citep{evaraw08}.

In novae, dust formation typically occurs between one and five months post-outburst and observations show that this timescale depends on the rate the nova fades in the optical (the speed class; \citealp{geh98}, \citealp{evaraw08}, \citealp{sha11}), with dust condensation time, $t_{\mathrm{cond}}$, being longer for slower novae. Indeed, Figure 11 in \citet{sha11} shows there appears to be a strong correlation between $t_{\mathrm{cond}}$ and the time it takes for the brightness of a nova to decline by two magnitudes, $t_{2}$. However this relationship has not been explained theoretically.

This work therefore aims to better understand the relationship between dust formation timescale and speed class. During a nova outburst the spectrum of the central hot source evolves as the temperature of the pseudo-photosphere increases (e.g.\ \citealp{bathar89}). At higher temperatures more radiation is systematically emitted at higher energies than the Lyman limit and therefore can be absorbed by neutral hydrogen (see Section~\ref{sec:lym}). Using this hypothesis, the basic theory described in Section \ref{simplistic} is modified and the results are compared to the observational data.

\section{Exploring the Speed Class - Formation Timescale Correlation}

\subsection{A simplistic model} \label{simplistic}

From energy balance considerations, \citet{evaraw08} find the dust formation timescale, $t_{\mathrm{cond}}$, is given by

\begin{equation} \label{eq:tcond}
t_{\mathrm{cond}} = \left[\frac{L}{16{\pi}V_{\mathrm{ej}}^2{\sigma}T_{\mathrm{cond}}^{4}}\frac{\langle Q_{\mathrm{a}}\rangle}{\langle Q_{\mathrm{e}}\rangle}\right]^{\frac{1}{2}},
\end{equation}

\noindent where  $L$ is the luminosity of the nova as seen by the grains (usually assumed to be the bolometric luminosity - see Section \ref{sec:lym}), $V_{\mathrm{ej}}$ is ejection velocity, $\langle Q_{\mathrm{a}}\rangle$ is Planck mean absorptivity, $\langle Q_{\mathrm{e}}\rangle$ is the Planck mean emissivity and $T_{\mathrm{cond}}$ is the dust condensation temperature. Hence, from Equation~(\ref{eq:tcond})

\begin{equation} \label{eq:tcondL}
t_{\mathrm{cond}}\propto L^{1/2}V_{\mathrm{ej}}^{-1}. 
\end{equation}

\noindent Using the Maximum Magnitude$-$Rate of Decline (MMRD) relationship from \citet{war08}, it can be shown that

\begin{equation}
2.5\log L\propto-\log t_{2}^{b_{2}} \nonumber
\end{equation}

\noindent and thus if $b_{2}\simeq2.5$ \citep{war08}, then

\begin{equation} \label{eq:lt2rel1}
L \appropto t_{2}^{-1}.
\end{equation}

\noindent An empirically determined relationship from \citet{war08} gives

\begin{equation}
\log V_{\mathrm{ej}}=3.57-0.5\log t_{2}, \nonumber
\end{equation}

\noindent where $V_{\mathrm{ej}}$ is in kms$^{-1}$ and $t_2$ is in units of days. In a survey of M31 novae, \citet{2011ApJ...734...12S} find a similar relationship from line widths. Therefore

\begin{equation} \label{eq:lt2rel2}
V_{\mathrm{ej}}\appropto t_{2}^{-0.5}.
\end{equation}

\noindent Substituting Equations~(\ref{eq:lt2rel1}) and (\ref{eq:lt2rel2}) into Equation~(\ref{eq:tcondL}) shows $t_{\mathrm{cond}}$ is then predicted to be approximately independent of $t_{2}$. Taking, for example, the value of $b_2 = 2.55\pm0.32$ from \citet{dowdue00} results in the $t_{\mathrm{2}}$ dependency of:

\begin{equation} 
t_{\mathrm{cond}}\propto t_2^{-0.01\pm0.06}. \nonumber
\end{equation}

\noindent However, as can be seen in \citet[their Figure 11]{sha11} this is clearly not the case and this basic analysis needs modifying. Our modification of the theory is described in Section~\ref{sec:lym}. If there was no significant MMRD relationship as discussed by \citet{2011ApJ...735...94K}, $t_{\mathrm{cond}}$ would be dependent on $t_2$ and would be given by

\begin{equation} 
t_{\mathrm{cond}}\appropto t_2^{0.5}. \nonumber
\end{equation}

\noindent A full quantitative analysis then reveals however that even if we assume very low luminosities (which are more conducive to earlier dust formation), dust is predicted to form far later than is ever observed.

\subsection{The Effect of the Evolving Underlying Nova Spectrum on Dust Formation} \label{sec:lym}

As a first step in refining the simplistic model described above, we assume the grain nucleation sites only see emission at wavelengths longer than the Lyman limit (see e.g. \citealp{evaraw94}). Although simple uniform-chemistry models predict that the ejecta would be fully ionised before dust formation takes place (e.g. \citealp{1984MNRAS.209..945M}), it is well known that the chemistry leading to the formation of nucleation sites needs dense, cool, neutral clumps (see \citealp{evaraw08} and references therein). Indeed, there is strong observational evidence -- in the form of Na\,{\sc i} (ionisation potential 5.14~eV) and CO emission (e.g.\ \citealp{1996MNRAS.282.1049E,2003ApJ...596.1229R,2009MNRAS.398..375D,raj12}) shortly before dust formation -- for the presence of cool hydrogen-neutral clumps which are likely at the outer edge of the ionised wind; nucleation sites in such clumps will indeed be exposed only to radiation longward of the Lyman limit. In our analysis we assume the dust is formed at the outer extremities of the ejected shell.

The initial step is to take the bolometric luminosity as constant (e.g.\ \citealp{war08}) and defined by the speed class and then we find the fraction of this luminosity that is red-ward of the Lyman limit for any given nova at any given time. We define the peak absolute magnitude, $M_{V}$, of a given nova using the MMRD relation

\begin{equation}
M_{V}=2.5\log t_{2}-11 \nonumber
\end{equation}

\noindent (see \citealp{war08}, and references therein). Assuming the bolometric correction BC = 0 at the peak of the visual light curve (as it seems most novae at maximum have an effective photospheric temperature of about 8000K - see \citealp{eva05}), $M_{V}$ is then converted to luminosity to give the bolometric luminosity, $L_{\mathrm{bol}}$, corresponding to each $t_{2}$.

The temperature of the pseudo-photosphere is at a minimum at the time of visual maximum, which is in turn reached within the first few days of outburst in most novae. At this time, almost all of the radiation is at wavelengths longer than the Lyman limit. The fraction of radiation red-ward of the Lyman limit can be calculated from the effective temperature, $T_{\mathrm{eff}}$, given by the following equation from \citet{bathar89}:

\begin{equation}
T_{\mathrm{eff}}=T_{0}\times 10^{{\Delta}V/2.5}, \nonumber
\end{equation} 

\noindent where $T_{0}$ is the pseudo-photospheric temperature at visual peak ($=8000$~K as noted above, \citealp{eva05}) and $\Delta V$ is the change in magnitude from peak. In order to find $T_{\mathrm{eff}}$ as a function of time we produced model optical light curves for each $t_{2}$ value.

To do this, first the outburst amplitude, $A$, for each $t_{2}$ value was estimated using the $57^{\circ}$ line in the amplitude$-$log $t_{2}$ relationship plot in \citet[the average flux observed from accretion disks with a random distribution of inclinations will be equal to that of $57^{\circ}$ system]{war95}. A standard exponential decline was then assumed. The time after outburst corresponding to each $\Delta V$ was calculated using

\begin{equation}
t = \frac{\ln A-\ln(A-\Delta V)}{\ln A-\ln(A-2)}t_{2}. \nonumber
\end{equation}

\noindent We also performed the analysis using a very simple optical decline of

\begin{equation}
\Delta V = \frac{2t}{t_2}, \nonumber
\end{equation}

\noindent which had very little effect on the final results.

\begin{figure}[ht]
\begin{center}
\includegraphics[scale=0.21]{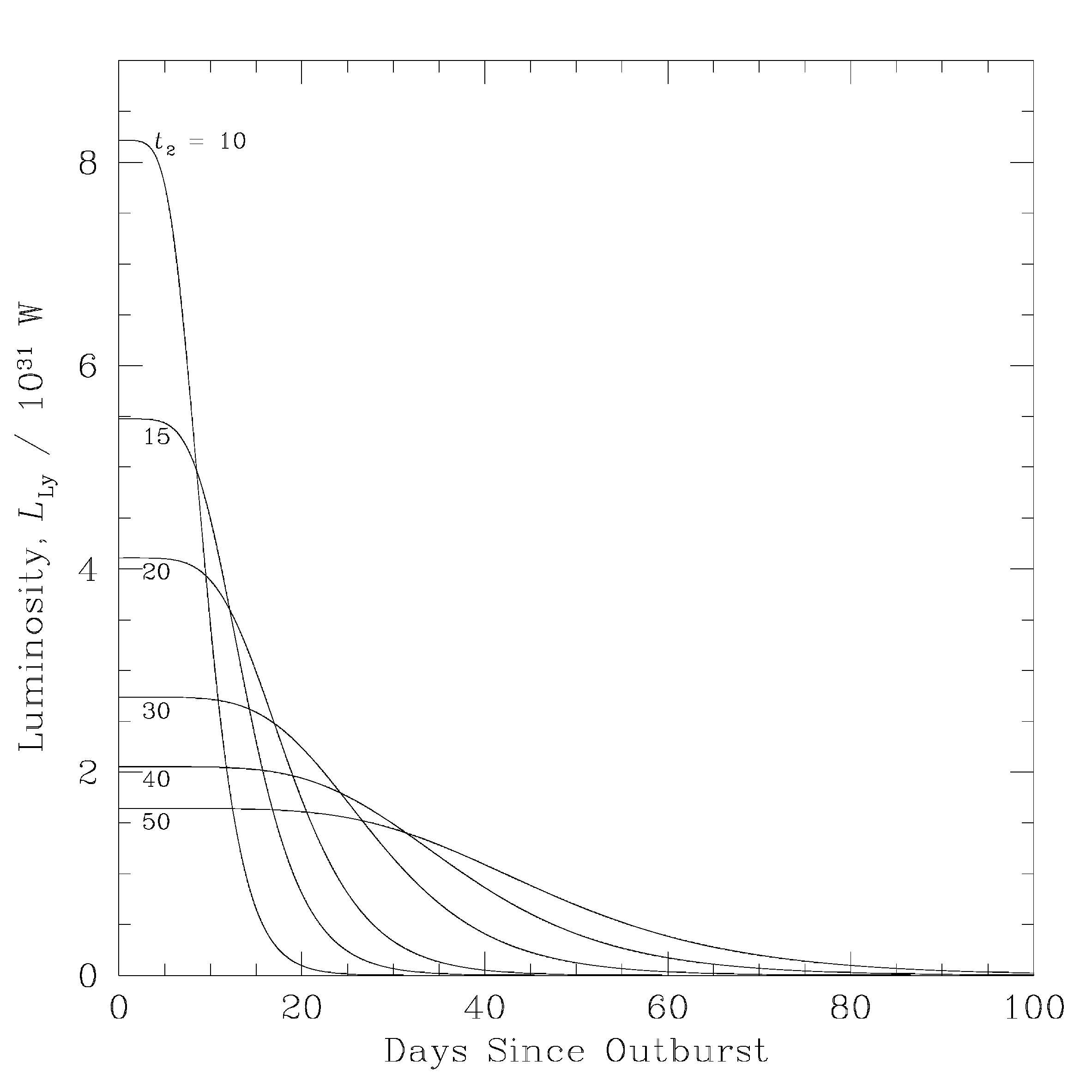}
\includegraphics[scale=0.21]{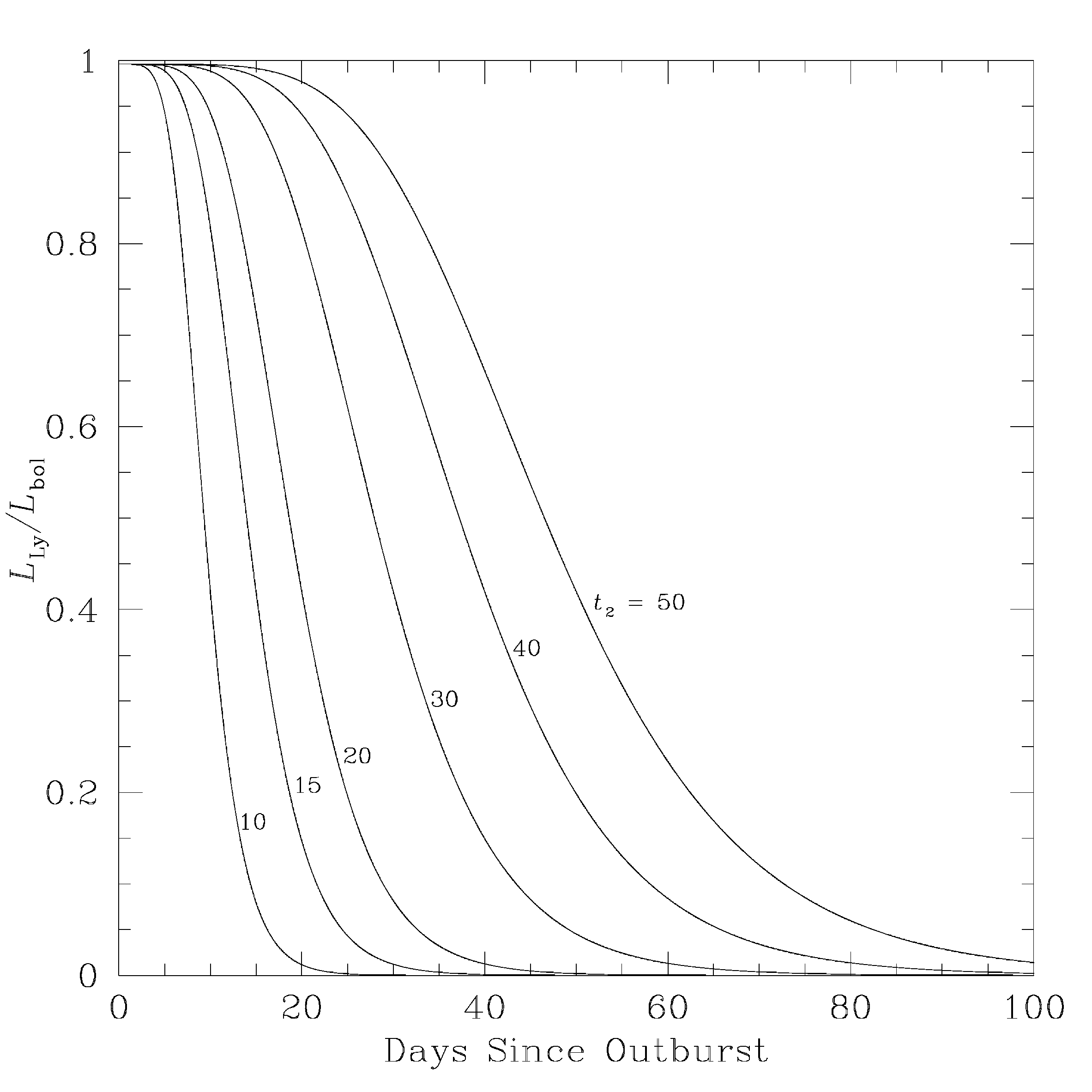}
\caption{How the luminosity seen by potential dust forming sites changes over time for different speed classes. The top plot shows the luminosity produced by a nova that will be seen by the grains for $t_2 = 10$~days (highest initial luminosity), $t_2 = 15, 20, 30, 40$ and $t_2 = 50$~days (lowest initial luminosity). The bottom plot shows the same, but expressed as a ratio of the total luminosity produced by a nova with a given $t_2$ value, with $t_2 = 10$~days being the fastest declining and $t_2 = 50$~days being the slowest.{\label{fig:luminosity}}}
\end{center}
\end{figure}

We then calculated how the luminosity red-ward of the Lyman limit, $L_{\mathrm{Ly}}$, declined over time for each $t_{2}$ value. Assuming the nova emits as a black body (although this is a first approximation for our purposes - see \citealp{hau08}), we integrated the black body function to find the luminosity red-ward of the Lyman limit that is received by the nucleation sites at a given time, $L_{\mathrm{Ly}}$.

\begin{equation}
L_{\mathrm{Ly}}=4{\pi}^{2}R^{2}{\int}_{91.2\;\mathrm{nm}}^{\infty}B_{\lambda}(T)d\lambda, \nonumber
\end{equation}

\noindent where $R$ is radius of the pseudo-photosphere. Figure \ref{fig:luminosity} shows $L_{\mathrm{Ly}}$ against $t$ for various $t_{2}$ values. It can be seen from the figure that, as expected, $L_{\mathrm{Ly}}$ drops much faster for faster novae. For example, potential dust formation sites in a $t_{2}=40$ days nova see more central source luminosity at $t=30$ days than potential sites in a $t_{2}=25$ days nova do at the same epoch, despite the constant bolometric luminosity being higher for the faster nova. 

As part of the revised theoretical exploration, we have considered different types of grain material, for example using data from \cite{zub96}. In terms of the size of nucleation centres we assume that C$_8$ as a nucleation centre acts as a solid sphere, where grain radius, $a\sim0.26$~nm (see \citealp{evaraw08}, and references therein). A Mie theory code was run to generate $Q_{\mathrm{abs}}$ values for a range of grain sizes from $0.26\le a\le 5$~nm over $0.05\le\lambda\le 1\;\mu$m for a condensation temperature, $T_{\mathrm{cond}}$, of 1200~K (appropriate for graphite in a 1:1 ratio C:O environment; \citealp{evaraw08}). Absorption efficiencies at longer wavelengths were found by extrapolation.

We chose to explore results for graphite and ACH2 (amorphous carbon is thought to form in H-rich environments, \citealp{zub96}), although the formation of spherical nucleation centres is of course less likely in the case of graphite. The results showed that, as expected in the Rayleigh regime, $Q_{\mathrm{abs}} \propto a$. We then calculated the Planck mean emission for graphite as

\begin{equation}
\langle Q_{\mathrm{e}}\rangle\simeq0.15aT_{\mathrm{d}}^{1.5} \nonumber
\end{equation}

\noindent and that for ACH2 as

\begin{equation}
\langle Q_{\mathrm{e}}\rangle\simeq 400aT_{\mathrm{d}}^{0.46}, \nonumber
\end{equation}

\noindent where $T_{\mathrm{d}}$ is the temperature of the grains (in Kelvin) and $a$ is in cm in each case. These Planck means are not valid over all temperatures, but sufficient for these calculations (i.e. they easily cover the range around 1200~K - the Planck mean for graphite is valid over the temperature range of approximately 500~K to 4000~K, with that of ACH2 being valid approximately between 500~K and 2000~K).

Thus, considering the energy balance between absorbed and emitted energy by a nucleation site for graphite
{\small
\begin{equation}
T_{\mathrm{d}}=\left[{\frac{5L_{\mathrm{bol}}}{12a{\sigma}^{2}T_{\mathrm{eff}}^{4}V_{\mathrm{ej}}^{2}t^{2}}}{\int}_{91.2\;\mathrm{nm}}^{\infty}B_{\lambda}(T_{\mathrm{eff}})Q_{\mathrm{abs}}(a,\lambda)d{\lambda}\right]^{0.18} \label{equ:graphite}
\end{equation}
}
\noindent and for ACH2
{\small
\begin{equation}
T_{\mathrm{d}}=\left[{\frac{L_{\mathrm{bol}}}{6400a{\sigma}^{2}T_{\mathrm{eff}}^{4}V_{\mathrm{ej}}^{2}t^{2}}}{\int}_{91.2\;\mathrm{nm}}^{\infty}B_{\lambda}(T_{\mathrm{eff}})Q_{\mathrm{abs}}(a,\lambda)d{\lambda}\right]^{0.22}. \label{equ:ACH2}
\end{equation}
}
\noindent We solved Equation (\ref{equ:ACH2}) numerically for dust condensation temperatures in the range of 1000~K to 1400~K and solved Equation (\ref{equ:graphite}) numerically for $T_{\mathrm{cond}}=1200$~K for reference.

\section{Results and Discussion} \label{results}

\begin{figure*}[ht]
\begin{center}
\includegraphics[scale=0.8]{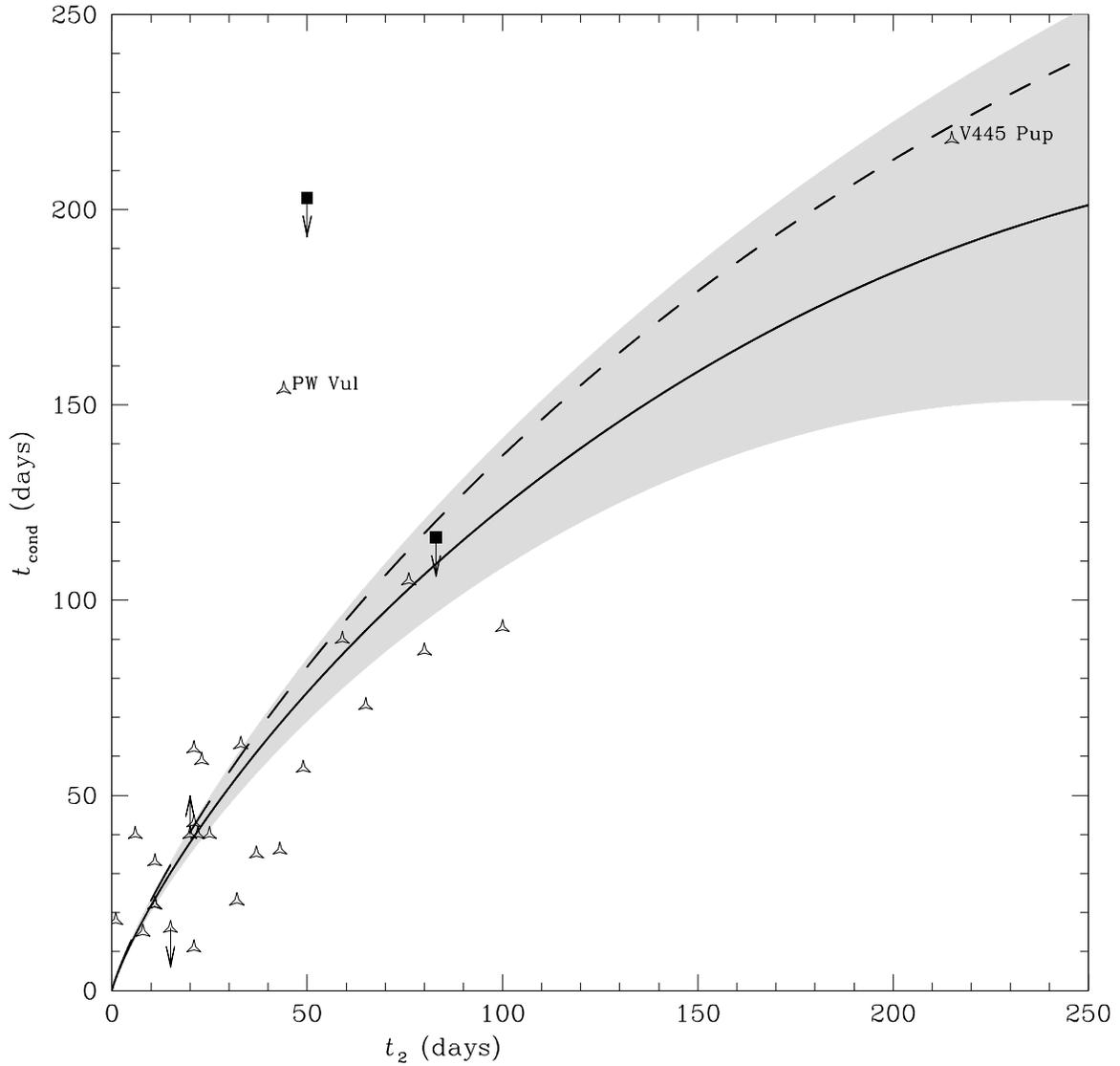}
\caption{Observational data from \citet{sha11} and references therein, with the additional data points listed in the text, showing dust condensation time, $t_{\mathrm{cond}}$, against the speed of visual brightness decline, $t_{2}$, in novae. The Galactic novae are shown as open triangles, whilst the two suspected dust producing novae in M31 are shown as squares. The outlying Galactic novae PW~Vul and V445~Pup are discussed in more detail in \citet{sha11}. The dashed line shows the results for the model for graphite when $T_{\mathrm{cond}}=1200$~K. The solid line shows the results for the model for ACH2 when $T_{\mathrm{cond}}=1200$~K, with the shaded area showing the results for ACH2 for a $T_{\mathrm{cond}}$ in the range between 1000~K and 1400~K (with the higher $T_{\mathrm{cond}}$ being the lower limit of the shaded area).\label{fig:speeddust}}
\end{center}
\end{figure*}

We have compared the predicted $t_{\mathrm{cond}}-t_{2}$ relationships from the above model with the data produced by \citet{sha11} using their observations and those of references therein. This is shown in Figure~\ref{fig:speeddust}, with additional points added for V1425~Aql \citep{kam97}, V1280~Sco \citep{che08}, V5579~Sgr \citep{raj11}, V496~Sct \citep{raj12} and Nova Cep~2013 \citep{mun13}. The graphite and ACH2 relationships shown in the figure were produced by performing the numerical integration described at the end of Section \ref{sec:lym} over a range of $t_{2}$ values.

As can be seen from Figure \ref{fig:speeddust}, the results for both graphite and ACH2 appear to agree well with the general trend of $t_{\mathrm{cond}}$ with $t_2$ from the observations, despite the simplistic assumptions made. There do not appear to be great differences between the relationships for the two types of carbon. The shaded area of the figure displays the relationship for ACH2 for $T_{\mathrm{cond}}$ between 1000~K and 1400~K. This shows that the relationship holds well for a range of $T_{\mathrm{cond}}$. We note that many specific data points at the lower $t_2$ values lie outside our relationship, but this reflects the simplicity of our model and the variation in parameters, including specific dust grain types, expected for individual nova outbursts.

It can also be seen in Figure~\ref{fig:speeddust}, that there is a gap in observational data between 100 $\gtrsim$ $t_2$ $\gtrsim$ 200. We note of course that the extreme case, V445 Pup, is an unusual helium nova. The apparent gap may however be due to the relative rarity of novae of these slow speed classes, and furthermore not every nova being observed sufficiently systematically at such late times to detect the epoch of dust formation. Overall, our results can be used to predict when dust will form in a nova of a given speed class and hence when, for example, infrared observations should be taken to detect the onset of dust formation.

The only known dust-forming nova where carbon-based dust has not been observed is QU~Vul, where SiO$_2$ dust was formed ($t_{2}=20$ days and $t_{\mathrm{cond}}=40$ days; \citealp{evaraw08,str10}), although this nova still appears to match our theoretical relationships for carbon-based dust. Several novae have been observed to form both carbon-rich and oxygen-rich dust \citep{evaraw08}, although the dust in CO novae is thought to consist mainly of amorphous carbon grains \citep{geh08}. As a future step it would be feasible to conduct a similar analysis for silicates, although these form later than carbon-rich dust where both types are formed in the same nova \citep{geh08}. Similarly, the application of more realistic spectral energy distributions for the emission seen by the grains, and any effects of re-emission of photons within the clumps, plus the effects of non-spherical geometry (e.g.\ \citealp{che12}) should be explored.

\acknowledgments
SCW is supported by an STFC research studentship. AWS is grateful for financial support through NSF grant AST-1009566. We thank an anonymous referee for valuable comments on the initial submitted version of this paper.

\bibliographystyle{apj}

\end{document}